\pdfoutput=1
\documentclass[twocolumn,superscriptaddress,showpacs,showkeys]{revtex4}

\usepackage{bm}
\usepackage{amsthm, amssymb,amsmath}
\usepackage{graphicx}
\DeclareGraphicsExtensions{.jpg,.jpeg,.pdf,.png,.mps,.eps}

\newcommand{\EQL}{\begin{equation}\label}
\newcommand{\EQ}{\begin{equation}}
\newcommand{\EN}{\end{equation}}
\newcommand{\BFG}{\begin{figure}}
\newcommand{\EFG}{\end{figure}}
\newcommand{\ITM}{\begin{itemize}}
\newcommand{\ITN}{\end{itemize}}
\newcommand{\ENM}{\begin{enumerate}}
\newcommand{\EEN}{\end{enumerate}}
\newcommand{\BEA}{\[\begin{array}}
\newcommand{\EEA}{\end{array}\]}
\newcommand{\EQA}{\begin{equation}\begin{array}}
\newcommand{\ENA}{\end{array}\end{equation}}

\newcommand{\obdot}[1]{\overset{\mbox{\boldmath$.$}}{#1}}

\newcommand{\bx}{\mbox{\boldmath$x$}}

\newcommand{\bomega}{\mbox{\boldmath$\omega$}}

\newcommand{\omhat}{\mbox{\boldmath$\hat{\omega}$}}

\newcommand{\half}{\mbox{$\frac{1}{2}$}}

\newcommand{\ominfty}{\mbox{$\|\bomega\|_\infty~$}}

\newcommand{\biband}{~and~}
\newcommand{\authone}[2]{#1 #2,}
\newcommand{\authtwo}[4]{#1 #2 and~#3 #4,}
\newcommand{\auththr}[6]{#1 #2, #3 #4 and~#5 #6,}

\newcommand{\authmanythr}[6]{#1 #2,~#3 #4,~#5 #6,~}

\newcommand{\private}[2]{ (private communication).}
\newcommand{\yjour}[6]{{ #6}{ #2}{ #3}{ (#1)}{ #4}{#5}.}

\newcommand{\yproc}[7]{ #7 in: {#4} (Eds.), #5, #6, #1, pp. #2{#3}.}

\begin{document}

\title{\Large 3D Euler about a 2D Symmetry Plane}

\author{Miguel D. Bustamante}
\email{mig_busta@yahoo.com}

\affiliation{Mathematics Institute, University of Warwick, Coventry CV4 7AL, United Kingdom}

\author{Robert M. Kerr}
\affiliation{Mathematics Institute, University of Warwick, Coventry CV4 7AL, United Kingdom}

\begin{abstract}
Initial results from new calculations of interacting anti-parallel Euler vortices are presented with the objective of understanding the origins
of singular scaling presented by Kerr (1993) and the lack thereof by Hou and Li (2006). Core profiles designed to reproduce the two results are
presented, new more robust analysis is proposed, and new criteria for when calculations should be terminated  are introduced and compared with
classical resolution studies and spectral convergence tests. Most of the analysis is on a $512\times 128 \times 2048$ mesh, with new analysis on
a just completed $1024\times 256\times 2048$ used to confirm trends.  One might hypothesize that there is a finite-time singularity with
enstrophy growth like $\Omega\sim (T_c-t)^{-\gamma_\Omega}$ and vorticity growth like $\ominfty\sim (T_c-t)^{-\gamma}$. The new analysis would
then support $\gamma_\Omega \approx 1/2$ and $\gamma>1$. These represent modifications of the conclusions of Kerr (1993). Issues that might arise
at higher resolution are discussed.

\end{abstract}




\maketitle


\section{\label{intro}Introduction}

One definition of solving Euler's three-dimension incompressible equations \cite{Euler1761} is determining whether or not they dynamically
generate a finite-time singularity if the initial conditions are smooth, in a bounded domain and have finite energy. The primary analytic
constraint that must be satisfied \cite{BKM84} is: \EQL{eq:BKM} \int_0^T\ominfty dt\rightarrow \infty \EN where \ominfty is the maximum of
vorticity over all space. To date, Kerr (1993) \cite{Kerr93} remains the only fully three-dimensional simulation of Euler's equations with
evidence for a singularity consistent with this and related constraints \cite{ConstFM96}. Growth of the enstrophy production and stretching along
the vorticity, plus collapse of positions, supported this claim \cite{Kerr93}. Additional weaker evidence related to blow-up in velocity and
collapsing scaling functions was presented later \cite{Kerr05}.

There is only weak numerical evidence supporting these claims \cite{Graueretal98,OrlandiCarnevale07}. In a recent paper, as described in one of
the invited talks of this symposium, Hou and Li (2006) \cite{HouLi06} found evidence that the above scenario failed at late times.

This contribution will first comment on four issues raised at the symposium, then present preliminary new results. The four issues are: \ITM\item
How should spurious high-wavenumber energy in spectral methods be suppressed?
\item What criteria should be used to determine when numerical errors
are substantial?
\item What effect do the initial conditions have on singular trends?
A cleaner initial condition is proposed.
\item We introduce a new approach for determining
whether there is singular behavior of the primary properties and the associated scaling. This is applied to both new and old data. \ITN

All calculations will be in the following domain: $L_x\times L_y\times L_z=4\pi\times4\pi\times2\pi$ with free-slip symmetries in $y$ and $z$ and
periodic in $x$ with up to $n_x\times n_y\times n_z=1024\times256\times2048$ mesh points. Using these symmetries only one-half of one of the
anti-parallel vortices needs to be simulated.

The ``symmetry'' plane will be defined as $xz$ free-slip symmetry through the maximum perturbation of the initial vortices and the ``dividing''
plane will be defined as the $xy$ free-slip symmetry between the vortices.

\section{How should spurious high-wavenumber energy in
spectral methods be suppressed?}

A generic difficulty in applying spectral methods to localized physical space phenomena is the accumulation of spurious high-wavenumber energy
that leads to numerical errors.

What is the best approach for eliminating these spurious modes? We have compared the old-fashioned 2/3rds dealiasing versus the recently proposed
36th-power hyperviscous filter \cite{HouLi06,HouLi07}.  Detailed tests to be described in a later paper show that the latter is better in the
sense that for several quantities, such as the peak vorticity, lower resolution calculations follow the high resolution cases longer.  But a
combination of the two approaches works even better, and that is what is used here.

Still, caution is required for any of these approaches as the hyperviscosity can dissipate small structures such as the anomalous negative
vorticity in the squared-off profile below.  Surprisingly the 36th-order hyperviscosity does not appear to produce the ghost vortices that are a
known artifact of lower-order schemes.

\section{What criteria should be used to determine when numerical errors
are substantial?} \label{sec:what criteria} There are traditionally two approaches to this problem, one emphasizing local quantities such as
\ominfty, and the other emphasizing global quantities such as the mean square vorticity or enstrophy.  We use both.

\subsubsection{Local quantities and resolution}

To determine local resolution it is important to check the convergence of local quantities such as: \ITM\item The maximum of vorticity
$\ominfty$. The location of \ominfty will be defined as $\bx_\infty$.
\item The local stretching of vorticity
\EQL{eq:alpha} \alpha=\omhat_i e_{ij}\omhat_j \EN where $\omhat=\bomega/|\omega|$ and $e_{ij}=\half(u_{i,j}+u_{j,i})$. \ITN

Following earlier work \cite{Kerr93,HouLi06}, we use the criteria that $\bx_\infty$ cannot be closer than 6 mesh points from the dividing plane.

\subsubsection{Integral quantities}

Examples of integral quantities we could monitor are: energy, circulation (which are in principle conserved), enstrophy and helicity (which are
in principle changing).

\ITM\item[i)] Energy is robustly conserved by spectral methods even when under-resolved and therefore is not a useful test. Convergence of the
energy spectrum \cite{HouLi06} is only a partial test because it neglects phase errors.

\item[ii)] Circulation in the upper half of the symmetry plane (i.e., the $z>0$ half of the $xz$-plane, which is perpendicular to the
primary direction of vorticity $y$) is conserved. Circulation in the equivalent half of the dividing plane is also conserved.  In all of the
initial conditions considered here, it is initially zero and ideally should remain so. Therefore, the circulations of the symmetry and dividing
planes, $\sigma_y = \int_{z>0} \omega_y(x,0,z,t) dx\,dz$ and $\sigma_z = \int_{y>0} \omega_z(x,y,0,t) dx\,dy$ were monitored.

We have found that serious depletion of $\sigma_y$ is controlled by $n_z$ and the time this begins is independent of the high-wavenumber filter.
Once $n_z$ is set, by convergence of \ominfty, we find that there is good convergence if $n_x=n_z/2$ and $n_y=n_z/4$. A later paper will provide
more details on these convergence tests. We will violate the condition on $n_y$ at late times due to current memory restrictions.

Without the circulation test, it is difficult to draw conclusions about the late times in Hou and Li (2006) \cite{HouLi06} where they claim to
see divergence from the scaling of Kerr (1993) \cite{Kerr93}.

\item[iii)] Enstrophy $\Omega$ grows in time, so one test is to check how it is balanced by its production $\Omega_p$, which we determine
 directly.
The enstrophy and its production are \EQL{eq:Omega} \Omega=\int dV \omega^2 \,,\qquad \Omega_p=2\int dV \omega_i e_{ij}\omega_j\,. \EN
\item[iv)] Helicity grows within the quadrant simulated (not over
the full anti-parallel geometry), but its production is determined by pressure which has not been calculated.

\ITN

\section{What is the effect of the initial conditions
on the potentially singular behavior?}

\subsection{Earlier descriptions}

Because ambiguities in the earlier description of the initial condition of \cite{Kerr93} led to differences in the initial condition of Hou and
Li (2006) \cite{HouLi06}, the community needs a clear description of a reproducible, clean initial condition that yields the trends of Kerr
(1993) \cite{Kerr93}.  Ideally, we want an initial condition whose vorticity is purely positive in the upper half of the symmetry plane, which
following Kelvin's theorem will
remain positive for all subsequent times. These steps were used \cite{Kerr93} to massage the vortex profile in order to achieve this:

\ITM\item[1)] The first step in creating the initial profile of the vorticity core is to use an explicit function where the value and all
derivatives went smoothly to zero at a given radius. See references in \cite{Kerr93} for earlier work that had used a similar profile. To this, a
localized perturbation in its position in $x$ was given \cite{KerrH89}.
\item[2)] The second step is to remove high-wavenumber noise by applying a
symmetric high-wavenumber filter of the form: $\exp(-a(k_x^2+k_y^2+k_z^2)^2)$. Kerr (1992) \cite{Kerr92} showed the undesirable side-effects if
this is not done. However, it has become apparent that the high-wavenumber filter is not sufficient. \ITN

\subsection{Effect of a negative region}
\begin{figure*}
\includegraphics[scale=0.9]{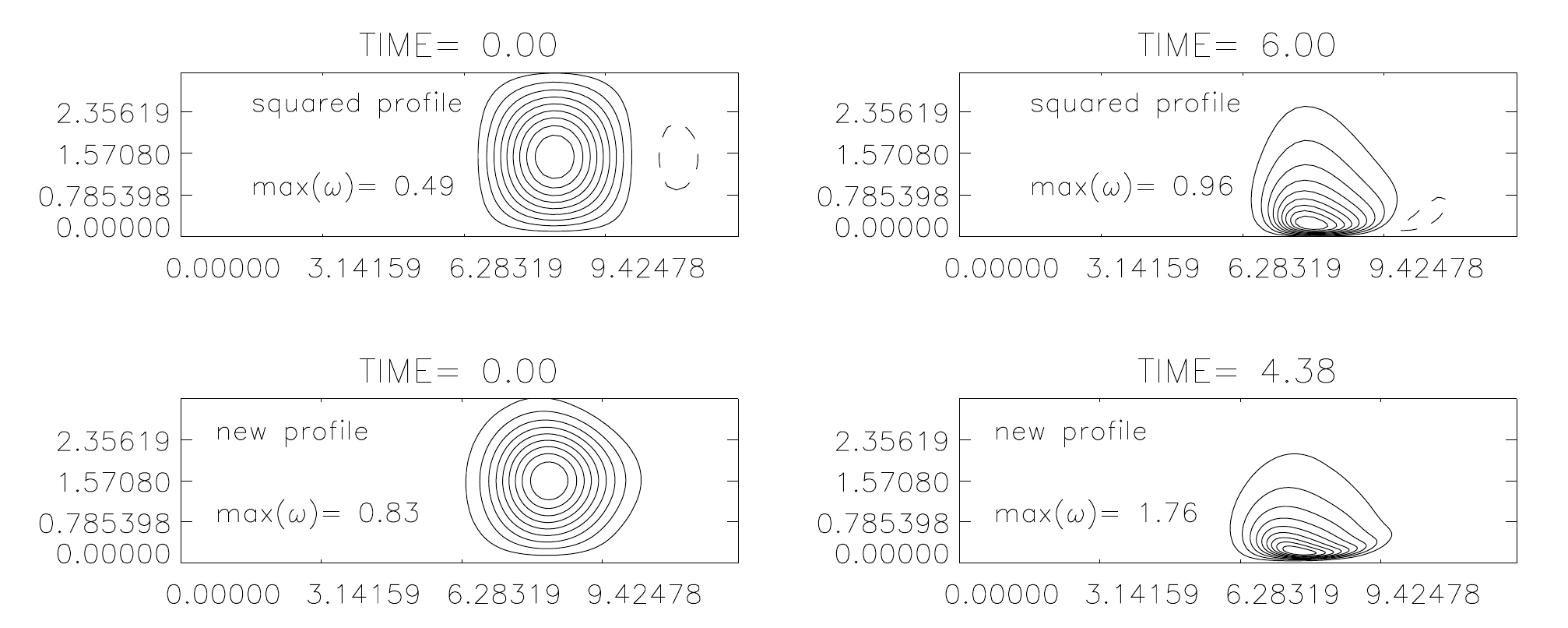}
\caption{$\omega_y$ in the symmetry plane from two initial conditions for $t=0$ and for early times with roughly the same growth in
$\max(\omega_y)$ in the symmetry plane. The first squared profile is nearly the same as used by Hou and Li (2006) \cite{HouLi06} using step 1)
with $\max(\omega_y)=0.49$ in the symmetry plane (over all space $\ominfty=\max(|\omega|)=.67$) and for step 2) a
 squared-off
high-wavenumber filter ($\exp(-a(k_x^4+k_y^4+k_z^4)$). Note the large negative vorticity in the lee (right) of the primary vortex as in Hou and
Li (2006) \cite{HouLi06} (Fig. 2) and how this is entrained underneath the primary vortex at $t=6$, whereupon the hyperviscous filter will
dissipate it. In the new profile step 3 is included: adding positive $\omega_y(z)$. In this case $\max(\omega_y)=0.83$ in the symmetry plane and
over all space $\ominfty=1.05$. There is no anomalous negative vorticity and numerical solutions require less resolution.} \label{fig:omy}
\end{figure*}

The upper frames in Fig. \ref{fig:omy} come from a reproduction of the squared-off profile of Hou and Li (2006) \cite{HouLi06} which follows the
procedure above with the exception of using a different high-wavenumber filter. Note the negative region in the lee of the primary vortex and how
this is sucked underneath the primary vortex at $t=6$. The $t=6$ frame represents the vortices simulated with the 36th-order hyperviscosity
\cite{HouLi06,HouLi07}. For $t>6$ this secondary vortex is dissipated by the hyperviscosity and circulation is dissipated, meaning these
calculations are not faithfully representing the Euler equations. Without the hyperviscosity, numerical noise would dominate as an extra boundary
layer needs to be resolved.

\subsection{Final step 3) for purely positive}

What is apparently missing from the previous description \cite{Kerr93} is the addition of a mean shear designed to remove the final negative
regions in the symmetry plane. This was achieved before \cite{Kerr93} as part of an interpolation procedure from a uniform mesh to a Chebyshev
mesh. Here it is imposed. Details will appear in a full paper. Initial vorticity in the symmetry plane and a slightly later time ($t=4.38$) are
shown. This is the initial condition for which we have now done up to $1024\times256\times2048$ calculations to assess the scaling proposed
earlier \cite{Kerr93}.

\section{A new approach for determining whether there is singular
behavior of the primary properties and the associated scaling.}

Once reliable data (according to the criteria discussed above) has been
 obtained,
it is common to interpret it in terms of power laws and other simple formulae.  For example, assume that \EQL{eq:fitfn} f(t)\sim
C/(T_c-t)^{\gamma} \,.\EN To properly find all three free parameters ($C$, $T_c$ and $\gamma$) to a set of points requires a minimization
procedure.

Kerr (1993) avoided this by assuming particularly simple values for $\gamma$ for several quantities.  In particular $\gamma=1$ was assumed for
\ominfty, for the maximum of the stretching of the vorticity \eqref{eq:alpha} in the symmetry plane: $\max(\alpha)|_{y=0}$, and for the enstrophy
production \eqref{eq:Omega}. This procedure was extended to the velocity by assuming that $\gamma_u=1/2$ for $\sup(|u|)$ \cite{Kerr05}.

While fits with these assumptions gave consistent results for the singular time $T_c$, this consistency existed only at late times when
resolution was becoming questionable. Analysis of this data by two new methods has shown that the lack of scaling at earlier times is due in part
to some of the more restrictive assumptions that were made.

\subsection{Three-parameter fitting}

Our first indication that earlier assumptions \cite{Kerr93,Kerr05} might be incorrect was obtained by allowing $\gamma$ to be free. The three
parameters ($C$, $T_c$ and $\gamma$) were then obtained as follows: by minimizing the sum of squares of the differences between the logarithm of
the data and the logarithm of the fit function, with respect to $C$ and $\gamma$, allows one to solve for these two parameters in terms of $T_c$.
Then the sum of squares is further minimized with respect to $T_c$ to obtain all three parameters.

This analysis was applied to \ominfty and $\Omega_p$, the enstrophy production, both of which previously were assumed to have $1/(T_c-t)$
behavior.

It was immediately observed that \ITM\item The fitting parameters depended upon the range of times
 chosen.
\item $\gamma$ in each case was consistently greater than 1.
\ITN

This would not be inconsistent with known bounds. Recall if power law behavior is expected for \ominfty, \eqref{eq:BKM} only requires that
$\gamma\geq 1$.

\begin{figure*}
\includegraphics[width=8.3cm,angle=0]{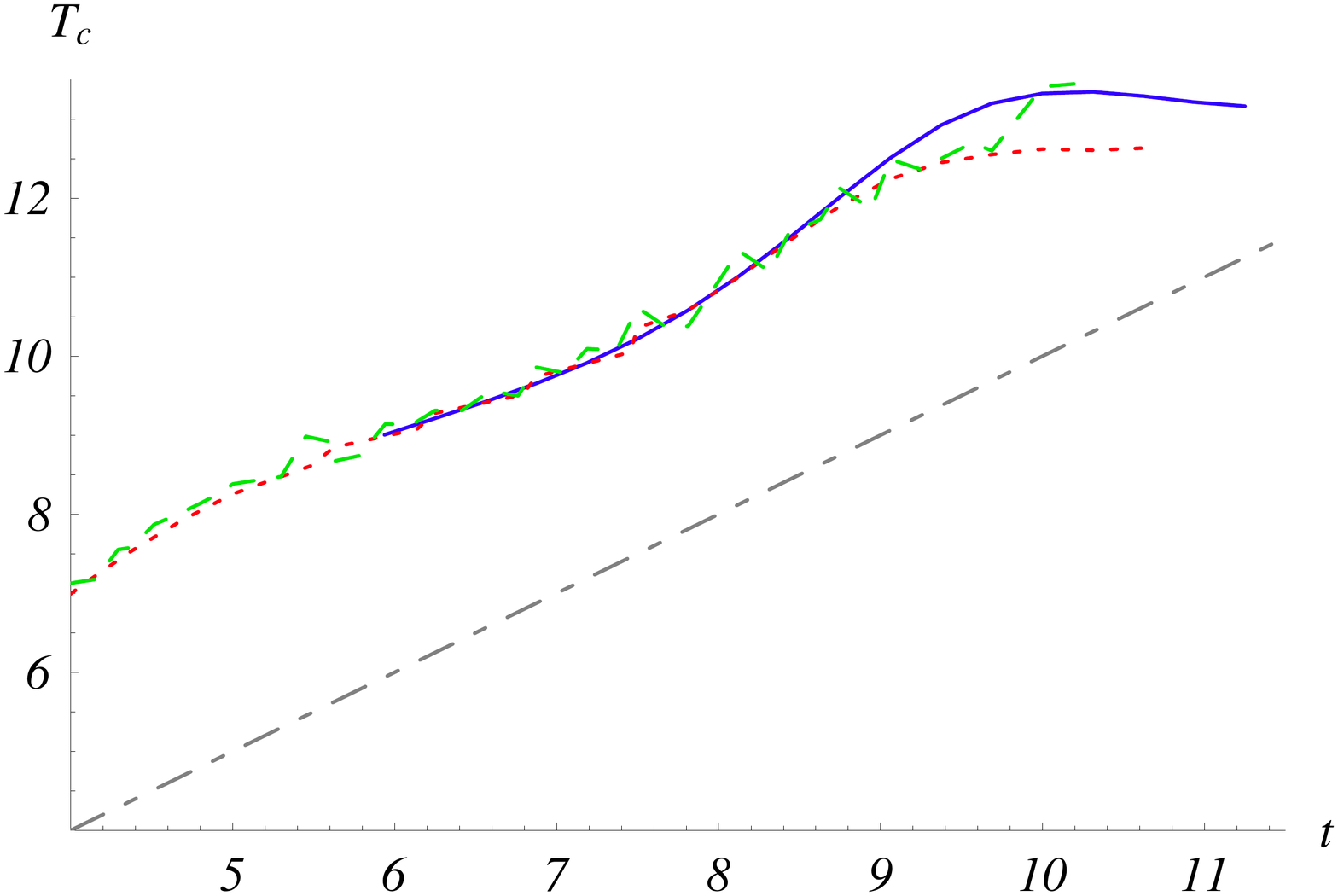}
\includegraphics[width=8.3cm,angle=0]{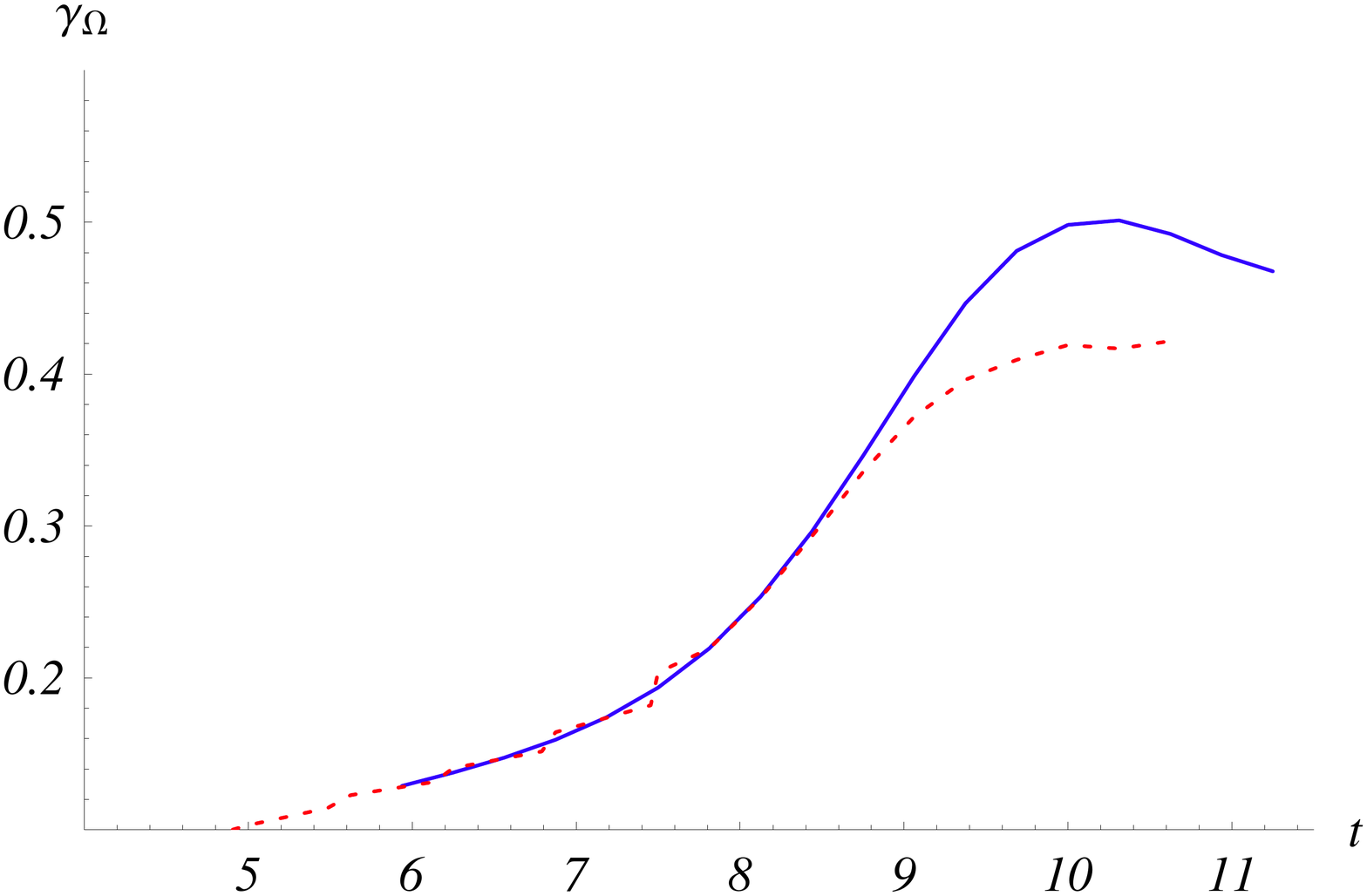}
\includegraphics[width=8.3cm,angle=0]{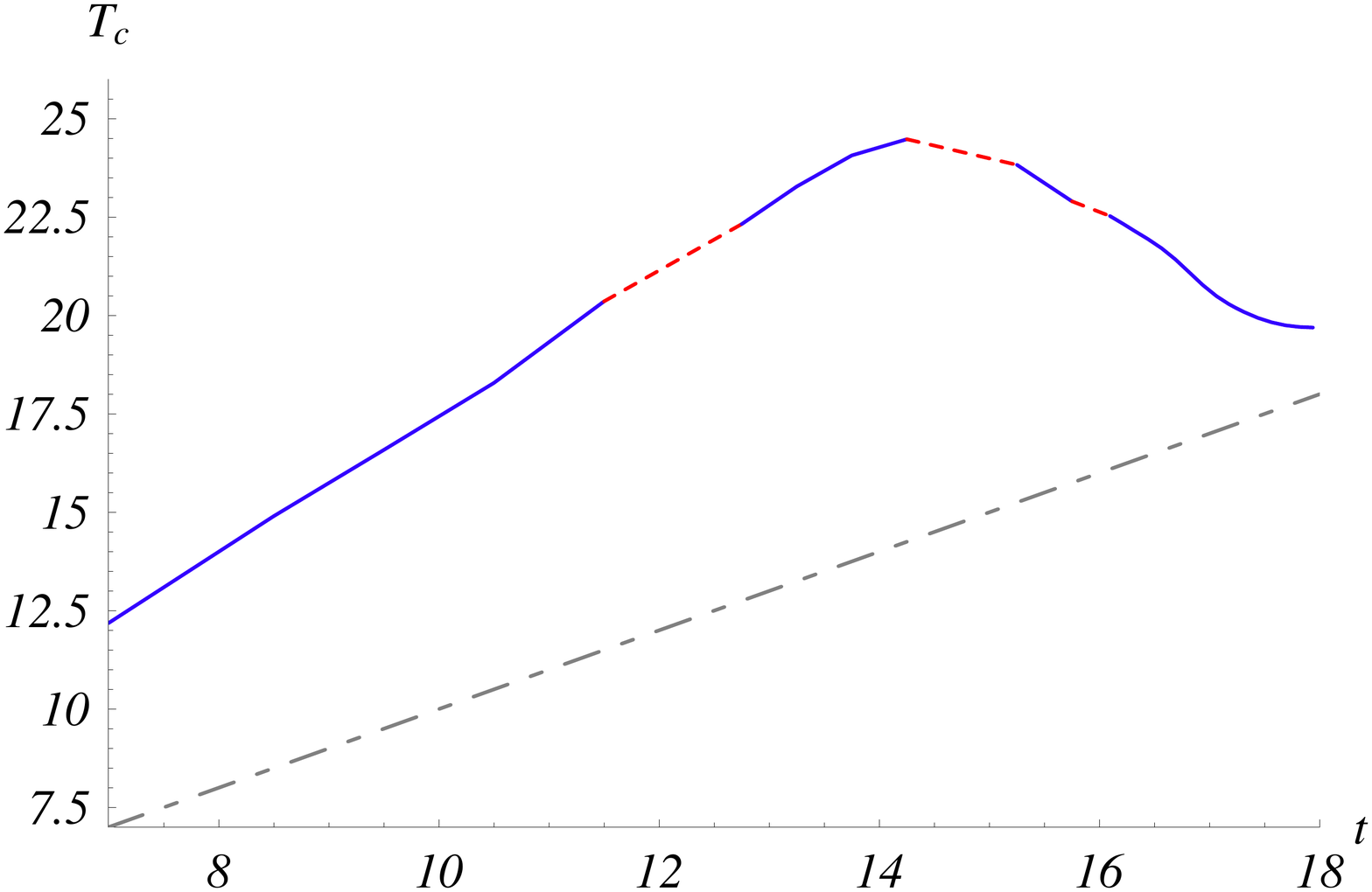}
\includegraphics[width=8.3cm,angle=0]{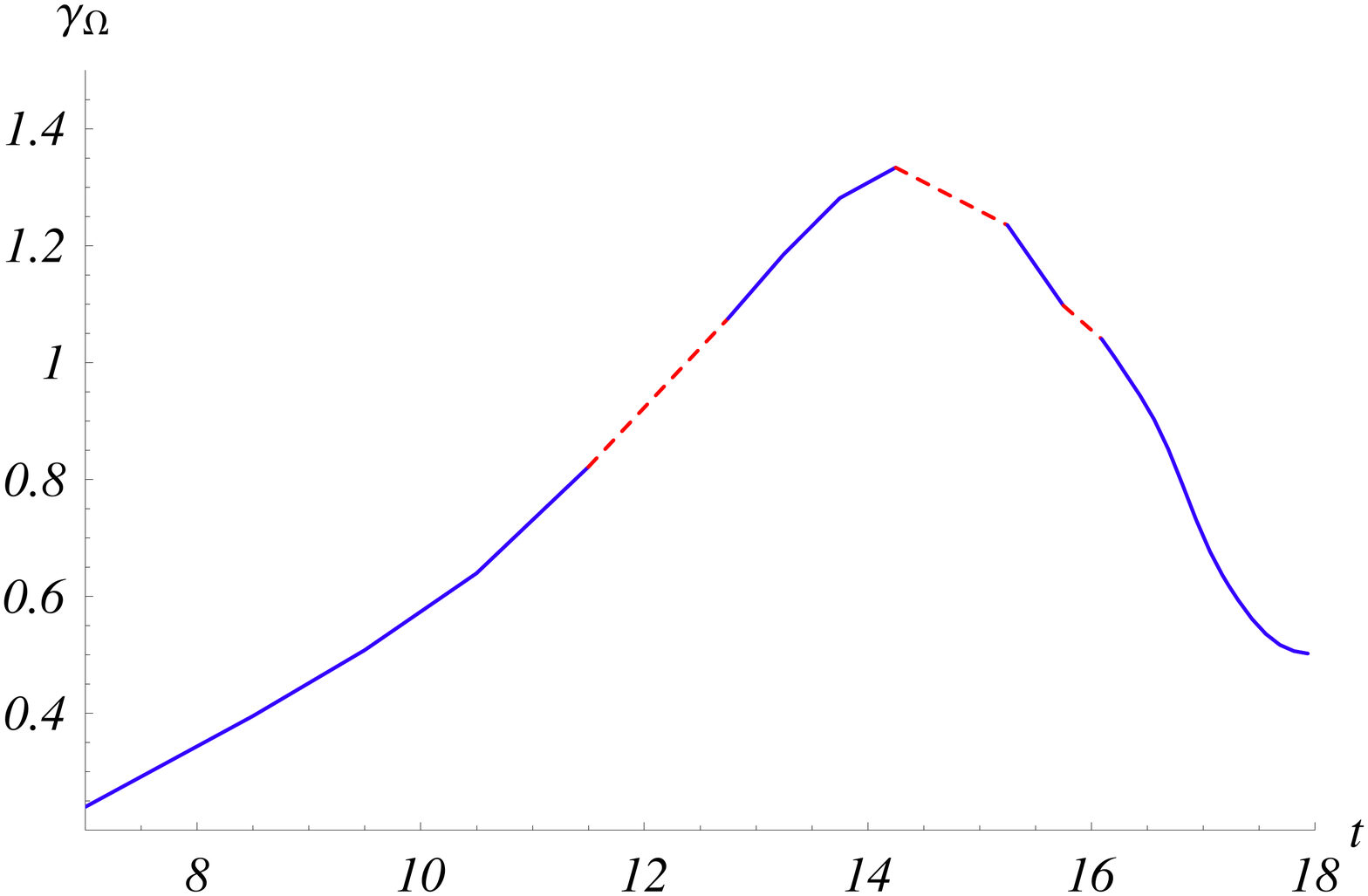}
\vspace{-0mm} \caption{Upper frames: Resolution study of the predicted singular time $T_c$ and the predicted exponent $\gamma_\Omega$ in the
power-law behavior of the total enstrophy $\Omega$ using the new data at resolutions: $512\times64\times2048$ (dashed), $512\times128\times2048$
(dotted)
and $1024\times256\times2048$ (solid). \\
Lower frames: Predicted $T_c$ and $\gamma_\Omega$ for the Kerr (1993) data at the highest resolution (solid). Dashed lines denote gaps in data.
In the graphs for the predicted $T_c$, the dash-dotted diagonal lines denote the $T_c=t$ singularity asymptote.} \label{fig:gammaFit and TcFit}
\end{figure*}


\subsection{Logarithmic time derivatives of enstrophy and \ominfty:
instantaneous two-parameter fitting} \label{sec:2param} The original analysis would be possible if there is a secondary
 quantity which
must go as $1/(T_c-t)$ if the primary quantity obeys the power law
 $1/(T_c-t)^\gamma$.
An example of a secondary quantity of that sort is the logarithmic time derivative of the primary quantity, which can be computed if we know
independently the quantity $f(t)$ and its time derivative $\dot{f}(t)$.

Therefore we propose a new approach to inferring a singular time and identifying the scaling behaviour:

\ITM\item Find a quantity $f(t)$ whose growth and the growth of its time-derivative can be determined directly. Consider the new function:
\EQL{eq:fitlog} g(t) = \left(\frac{d}{dt}\log
 f(t)\right)^{-1}=\frac{f}{\obdot{f}} =\frac{1}{\gamma}(T_c-t)\,. \EN
Note that the parameter $C$ drops out and the function is linear, so we can predict instantaneous values of $\gamma$ and $T_c$ by fitting this
new function using adjacent points in time.
\item Calculating in this manner, using nearest neighbours in time,
yields instantaneous predicted singular times $T_c(t)$ and power laws $\gamma(t)$. These instantaneous parameters will generically depend on
time.
\item See if these converge or relax (as time increases).
\ITN

Pairs of quantities to which this procedure can be applied are: \ITM\item \ominfty and its logarithmic time derivative $\alpha_\infty$ (local
stretching at the point $\bx_\infty$).
\item Enstrophy $\Omega$ and its production $\Omega_p$ \eqref{eq:Omega}
where we assume that \EQL{eq:gOmega} \Omega\sim \frac{C_\Omega}{(T_c-t)^{\gamma_\Omega}}\,. \EN
\item Helicity in the simulated quadrant of space and its production.
\ITN

 This approach is applied to enstrophy and its production on our highest resolution simulations. Because enstrophy and its production are
global quantities they converge numerically longer (to $t=11.25$) than \ominfty, for which $\bx_\infty$ is less than 6 mesh points from the
dividing plane for $t>10$. Running estimates for $T_c$ and $\gamma_\Omega$ are shown in Fig. \ref{fig:gammaFit and TcFit}. For the new data
(upper frames), the latest data point gives an estimate $T_c \approx 13.16$ and $\gamma_\Omega \approx 0.47$. The bottom frames show the same
analysis applied to the data used by Kerr (1993) \cite{Kerr93}. Nearly identical power laws are obtained ($\gamma_\Omega \approx 0.50$), with a
predicted singular time $T_c\approx 19.69$, greater than in Kerr (1993).

One advantage of finding running estimates of $T_c$ is that it can be used to identify cases that are not singular, or would take an unusually
long time to become singular. This is done by looking at the instantaneous estimated value of $T_c$. If $T_c$ continues to increase with time,
then there is evidence for regularity. In both the new calculations and for the data from Kerr (1993), eventually the estimated $T_c$ decreases
and relaxes to a finite value. It is quite possible that there is a large pre-factor in front of the power law, which the time dependence of the
estimated $\gamma_\Omega$ and $C_\Omega$ might be able to shed light on.

This approach assumes smooth values for both quantities in a pair. Unfortunately, we have found that because \ominfty sits on a steep gradient of
$\alpha$, values of $\alpha_\infty$ on the lower resolution mesh were not smooth enough in time to perform this analysis. The analysis will be
attempted on the higher resolution (\ominfty,$\alpha_\infty$) data when that additional analysis of the new data sets is available.

 \subsection{Convergence studies: Is the evidence for singularity conclusive?}

Further tests at higher resolution are needed to support the singular trends seen here. Both current cases (old and new data) could be reliably
integrated up to times $t \approx T_c-2.75$. This would only be the beginning of the asymptotic regime of the potentially singular solution, as
is suggested by the late-time behavior of the curves for the predicted singular time $T_c$ in Fig. \ref{fig:gammaFit and TcFit}. New calculations
in progress should go beyond that barrier and help test the validity of the hypothesis of finite-time singularity.

In this subsection we show resolution studies with $n_z$ fixed to give a flavor of what will be shown in the next paper (we have also made
resolution checks with fixed $n_x$ or $n_y$ and varying the other two, not shown here). The resolution study is a classical tool to validate and
find reliability times for the numerical results. Another now widely accepted study that we present is a spectral convergence test (Sulem et al.,
1983 \cite{1983JCoPh..50..138S}) used recently by Cichowlas and Brachet (2004) (see \cite{Cich04} and references therein), where the exponential
decay of the energy spectrum as a function of the wavenumber is employed to give a criteria for the reliability time. Finally we complement the
above classical tests with our newly proposed tests of reliability: conservation of circulation through the symmetry plane and conservation of
circulation through the dividing plane.

We consider first the behavior of local quantities, in the sense of Section \ref{sec:what criteria}. Figure \ref{fig:maxOm and spec k_x} (top) is
a resolution study of the time dependence of the maximum vorticity; the bottom figure is a $t=10$ anisotropic energy spectrum $E(k_x,t)$, defined
by averaging the Fourier transform ${\bf \hat u}({\bf k',t})$ of the velocity field on flat duplicated sheets of width $\Delta k_x = 1$,
\begin{equation}\nonumber
E(k_x,t) = {\frac1 2} \sum_{k_x-\Delta k_x/2< |k'_x| <  k_x + \Delta k_x/2} |{\bf \hat u}({\bf k',t})|^2 \, .
\end{equation}
Following \cite{Cich04}, we fit: $\log E(k_x) = C - n \log(k_x) - 2 \delta \, k_x$ (solid line in the figure). The test consists in monitoring
the parameter $\delta$ as a function of time. The idea is that $\delta$, being the width in the complex plane of the analyticity strip of the
velocity field, should always be numerically resolved, at least by the mesh size. Another way to look at this condition is to ask that the
contribution of the exponential term to the change of $\log E(k_x)$ from the largest to the smallest scale allowed by the numerical resolution,
be greater than a prescribed factor. In more explicit terms, we can only fully trust the simulation up until the condition $\delta \,
k_{x}^{\max} \geq 1$ is violated, where $k_x^{\max} = n_x/3$ is the maximum relevant wavenumber of the Fourier representation. Notice that
different authors use different factors in the RHS of the last inequality. For our $t=10$ spectrum in resolution $1024\times256\times2048$ we
obtain $\delta \, k_x^{\max} \approx 1.07$, and therefore our simulation is validated by this method up to $t=10$. In this way we could
extrapolate the convergence of $\ominfty$ up to $t=10$, whereas a conservative extrapolation based solely on the resolution study in figure
\ref{fig:maxOm and spec k_x} (top) would see the $1024\times256\times2048$ computation converged up to $t=9$.

\begin{figure}
\vspace{-0mm}
\includegraphics[width=8.3cm,angle=0]{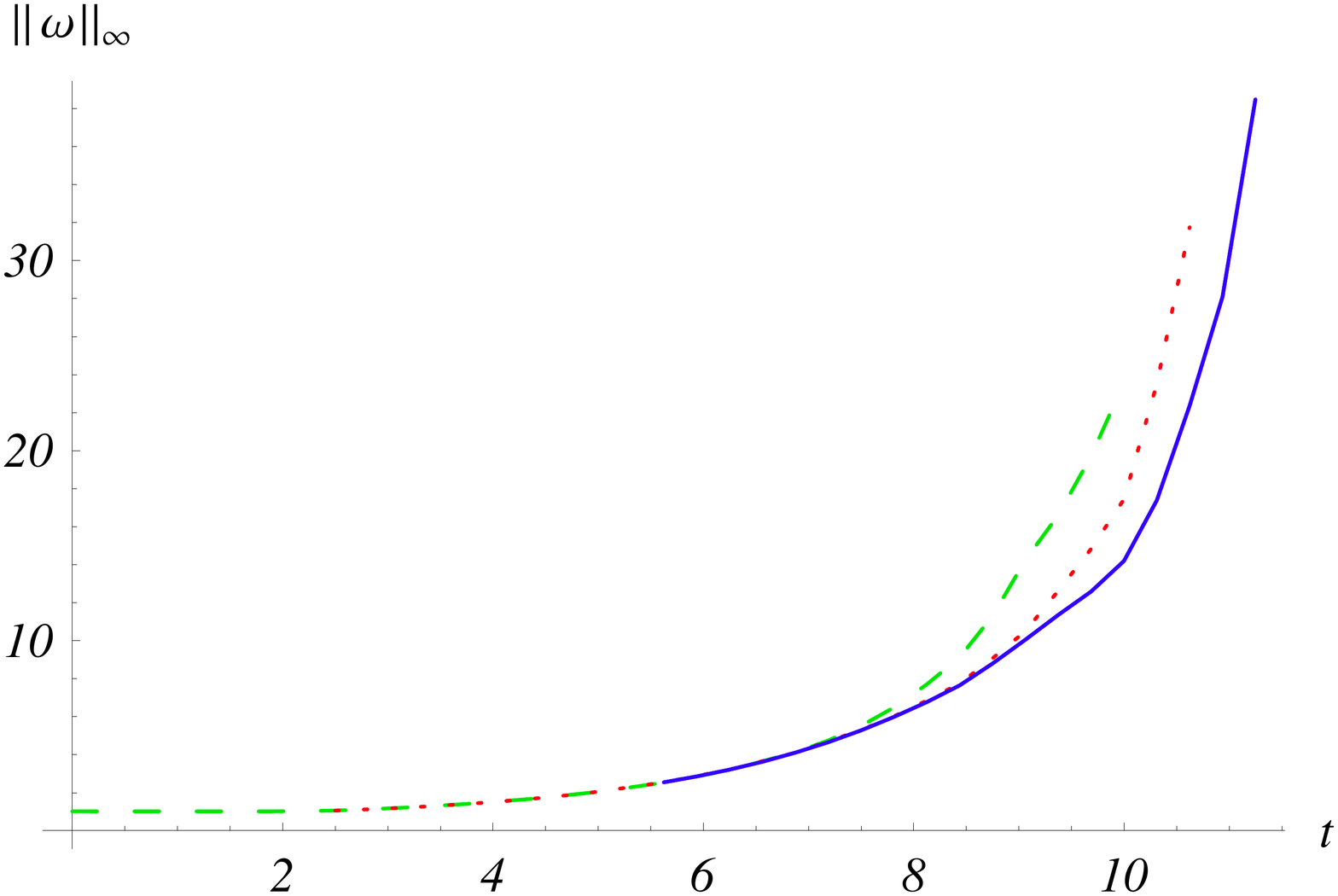}
\includegraphics[width=8.3cm,angle=0]{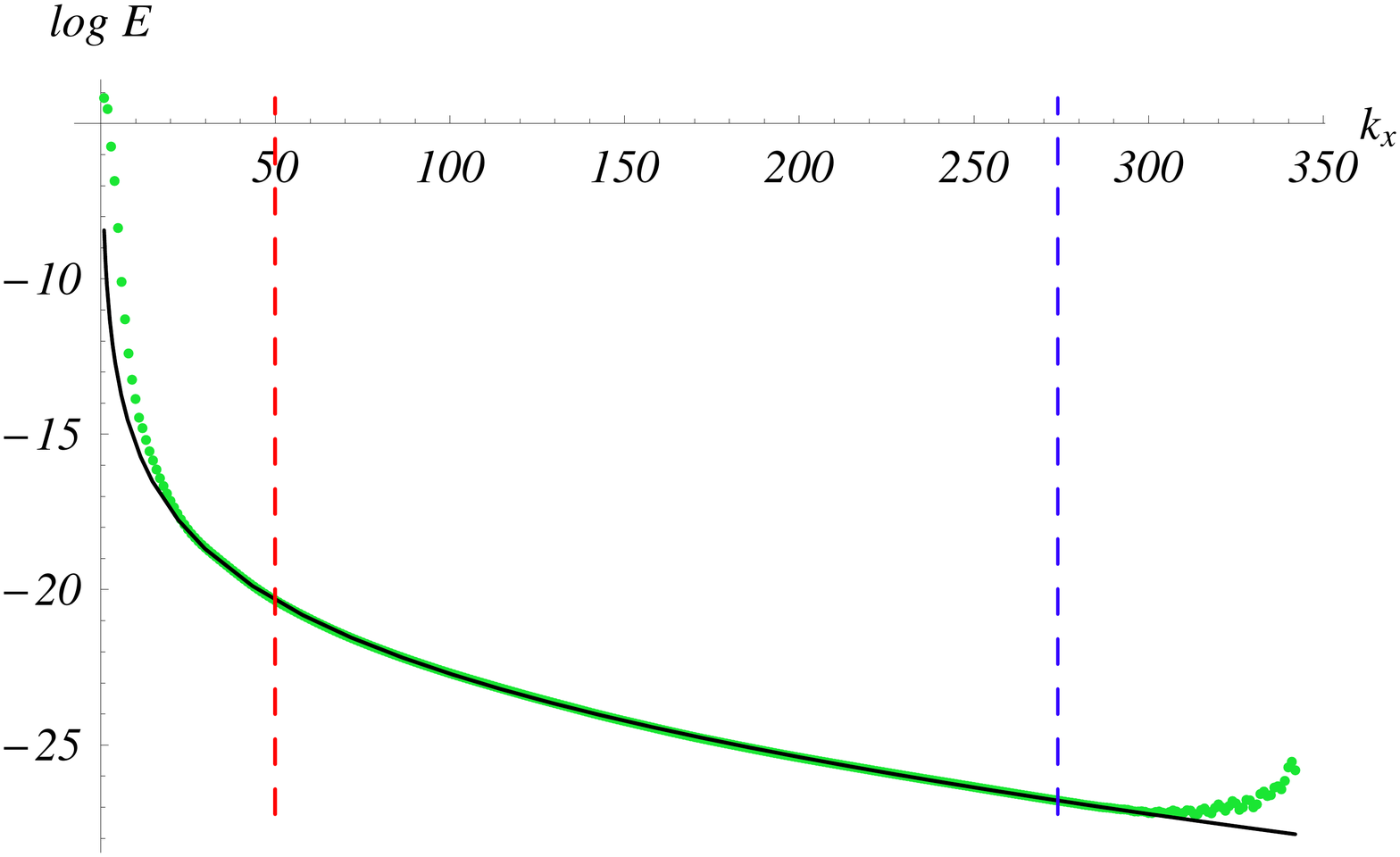}
\vspace{-0mm} \caption{Top: Resolution study of $||\omega||_{\infty}$, for resolutions $n_x \times n_y \times n_z$ of: $512\times64\times2048$
(dashed), $512\times128\times2048$ (dotted) and $1024\times256\times2048$ (solid). Bottom: Anisotropic energy spectrum (direction $k_x$) at time
$t=10$ for resolution $1024\times256\times2048$. Points correspond to numerical data. The solid curve corresponds to the fit of the spectrum
according to $\log E(k_x) = C - n \log(k_x) - 2 \delta \, k_x$, where the fit interval is defined by the vertical dashed lines.} \label{fig:maxOm
and spec k_x}
\end{figure}

\begin{figure}
\vspace{-0mm}
\includegraphics[width=8.3cm,angle=0]{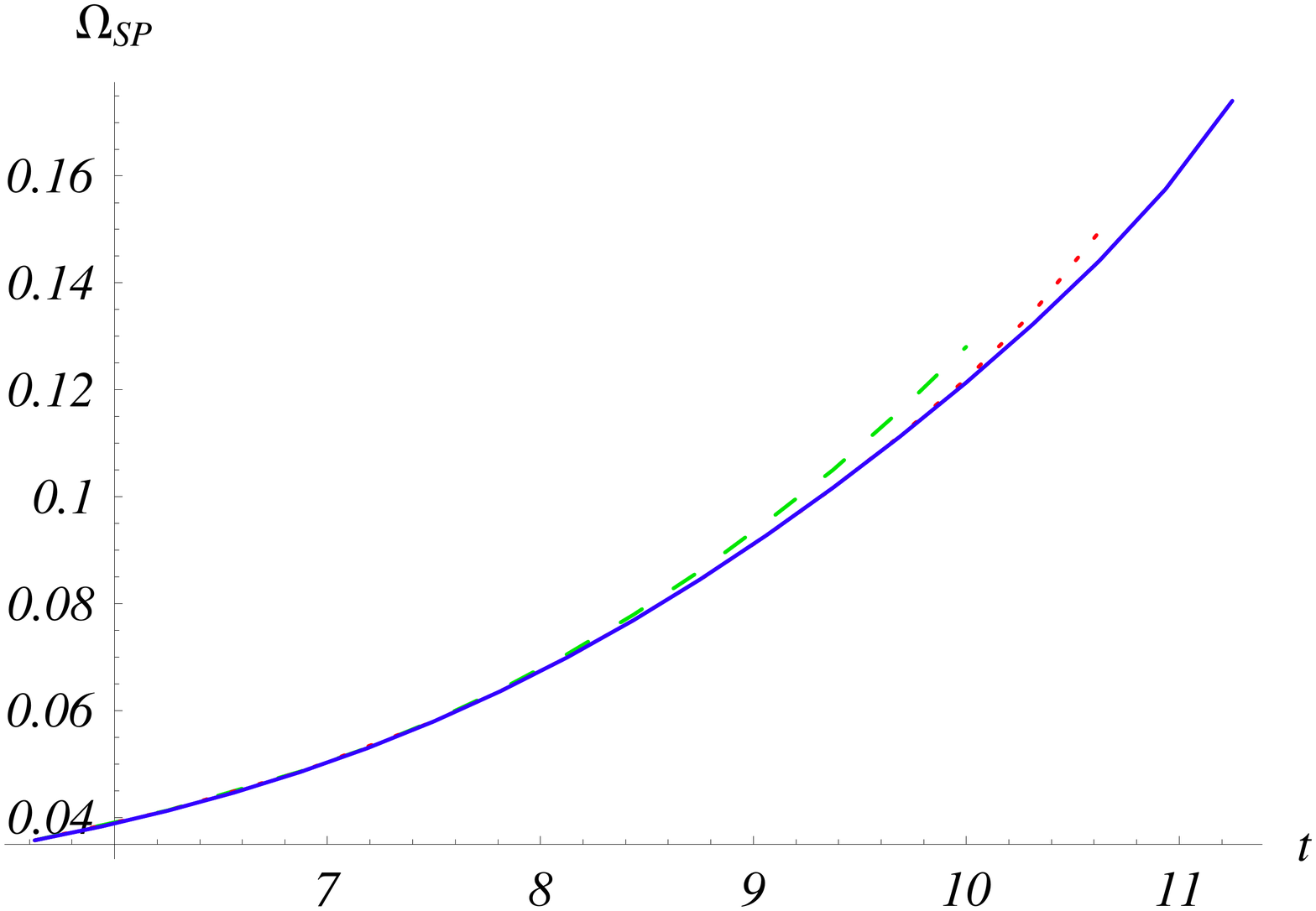}
\includegraphics[width=8.3cm,angle=0]{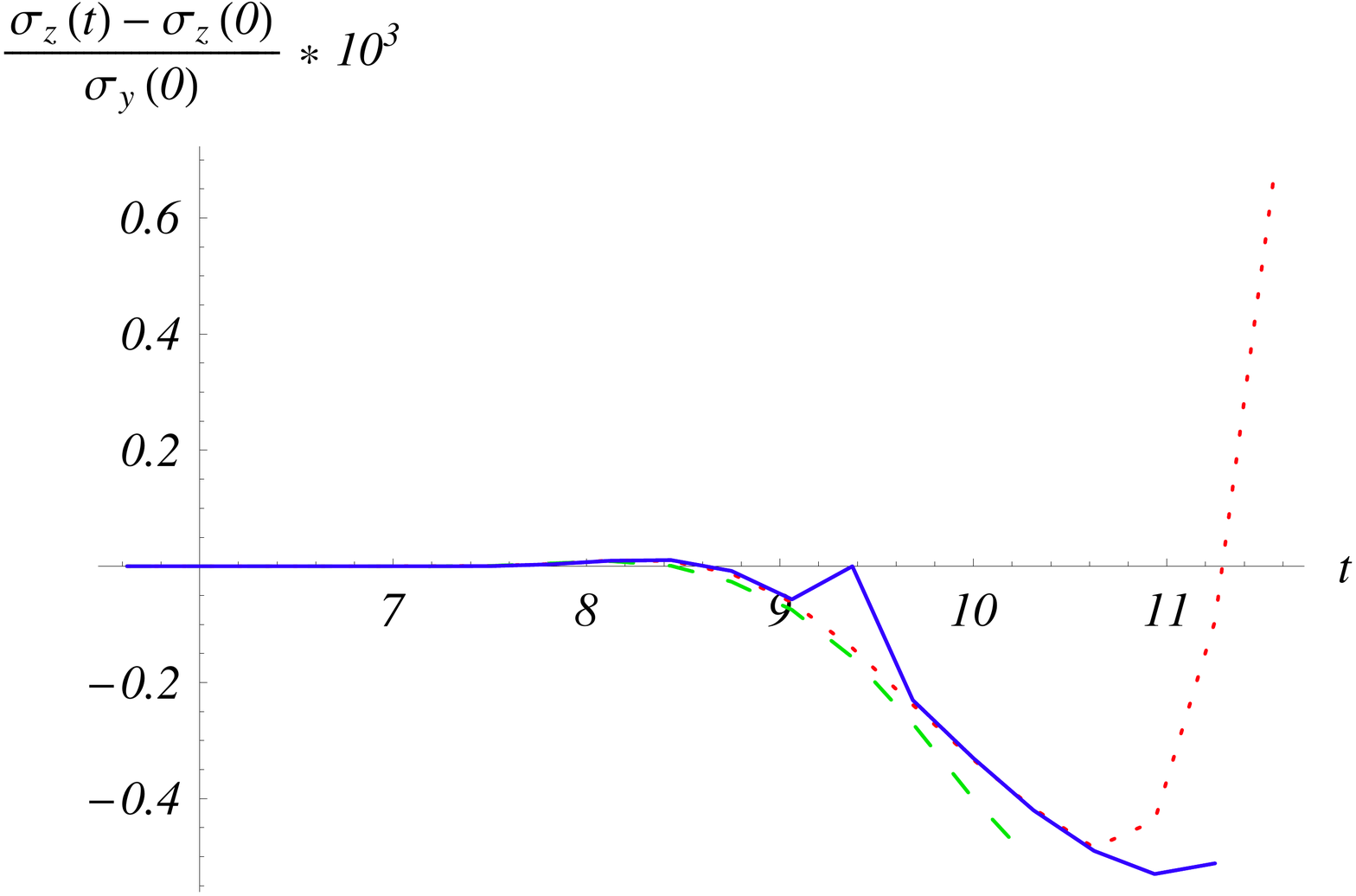}
\vspace{-0mm} \caption{Resolution study of: the enstrophy in the symmetry plane $\Omega_{SP}$ (top); the error in the circulation through the
dividing plane $\sigma_z$ normalized with the initial circulation through the symmetry plane $\sigma_y$ (bottom), for resolutions
$512\times64\times2048$ (dashed), $512\times128\times2048$ (dotted) and $1024\times256\times2048$ (solid).} \label{fig:enstrSP and errDP}
\end{figure}

We consider now the behavior of $2$D integral quantities. Due to its $2$D character, enstrophy in the symmetry plane $\Omega_{SP} = \int_{y=0}
\omega_y^2 \,dx \,dz$, shown in figure \ref{fig:enstrSP and errDP} (top), is a more sensitive measure than total enstrophy $\Omega$
(eq.(\ref{eq:Omega}), figure not shown), which converges more rapidly than $\Omega_{SP}$. A conservative extrapolation would imply convergence of
$\Omega_{SP}$ up to times $t \lessapprox 11$.

Figure \ref{fig:enstrSP and errDP} (bottom) is a resolution study of the normalized error in the conservation of circulation through the dividing
plane $\sigma_z$. We observe that, for a given resolution, the numerically induced deviation in $\sigma_z$ becomes unstable after a certain time.
Errors (and fluctuations thereof) less than $10^{-4}$ are acceptable, as long as they are stable. Then, a reasonable reliability time can be
defined for each resolution as the time when the error in $\sigma_z$ attains its last extremum before the instability takes over. According to
this criteria we conclude that the simulation at resolution $512\times128\times2048$ (dotted line) is converged up to $t \approx 10.7$ and the
simulation at resolution $1024\times256\times2048$ (solid line) converges up to $t \approx 11.25$. In order to display the unstable behavior of
the mid-resolution simulation (dotted line), we show data beyond its reliability time.

Finally we return to Fig. \ref{fig:gammaFit and TcFit}, considering all three resolutions.  For each resolution, $T_c$ has a peak for $t\approx
9-10$, and then asymptotes, well within the reliability time for the highest resolution. Similar trends towards convergence appear for
$\gamma_\Omega$. The key question regarding the existence of a finite-time singularity is if the curve for predicted $T_c$ crosses the asymptote
$T_c=t$ (dashed-dotted line) in a finite time or not, but we need further tests at higher resolutions (to be shown in a future paper) in order to
conclude on these matters.

We have enough resolution to conclude that the power laws are not the ones proposed in Kerr (1993,2005) \cite{{Kerr93},{Kerr05}} but not enough
resolution to reach definitive conclusions on the singular behavior, since $\ominfty$ does not converge as rapidly as the volumetric quantities
studied.


\begin{figure*}
\vspace{-70mm}
\includegraphics[width=16.0cm,angle=0]{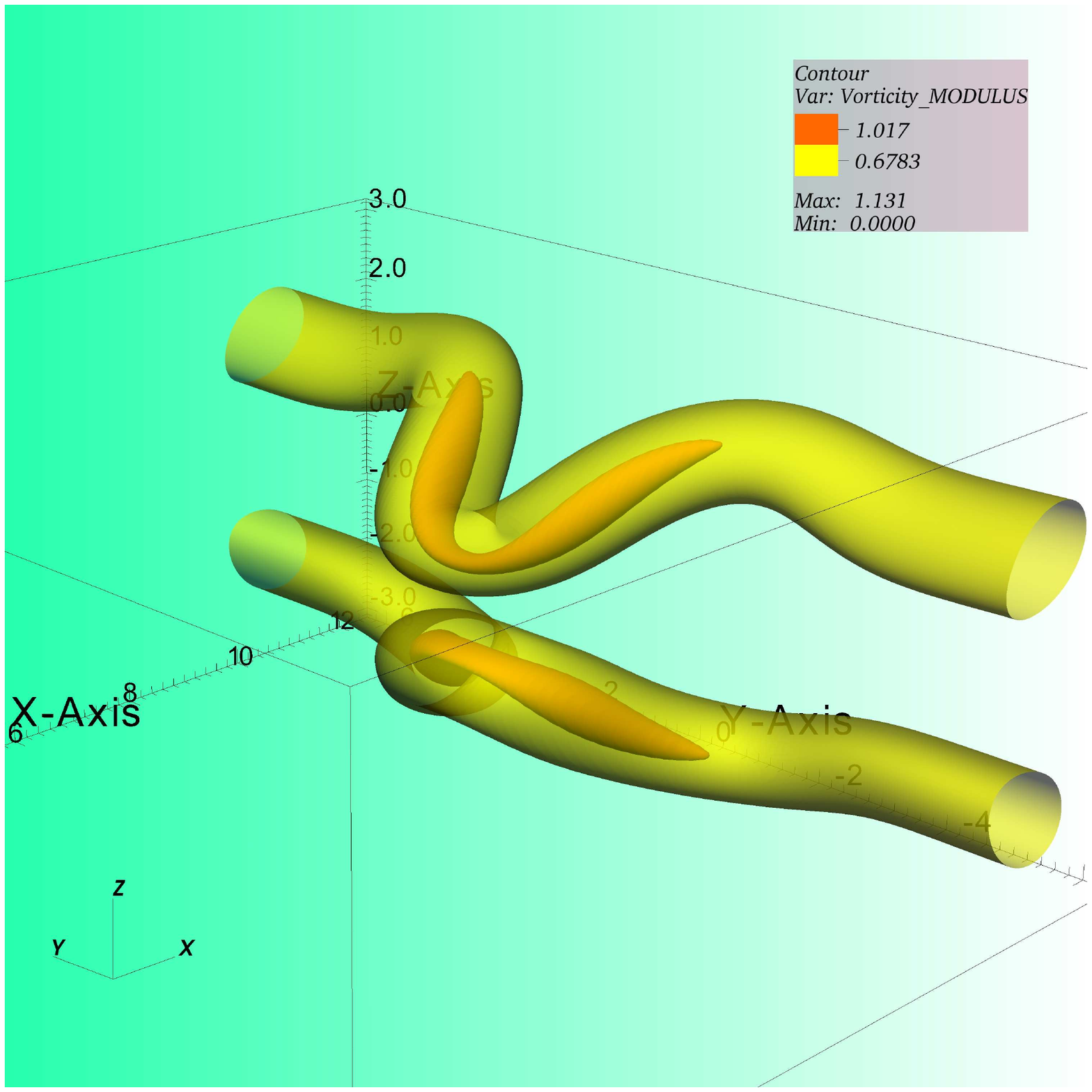}
\vspace{-0mm} \caption{Euler anti-parallel vortices in full periodic domain near $t=2.51.$ Bright (yellow online) tubes are isosurface contours
of vorticity modulus corresponding to $60\%$ of the instantaneous maximum of vorticity modulus. Dark (red online) elongated blobs are isosurfaces
corresponding to $90\%$ of the maximum of vorticity modulus.} \label{fig:v71c}
\end{figure*}

\section{Graphics}

In this section, 3D isosurface contours of the vorticity modulus are shown, corresponding to the simulation of initially anti-parallel vortices,
using a resolution of $512 \times 128 \times 2048$ in the fundamental quarter of the full domain, corresponding to an effective resolution of
$512 \times 256 \times 4096$ in the full domain. For memory-optimizing purposes, the output data used to make the figures has an effective
resolution of $512 \times 128 \times 1024$, corresponding to a memory size of $65$ MB. The freeware visualization program VisIt has been used to
make the plots.

Fig. \ref{fig:v71c} shows the vortices after some time of evolution in the whole periodic domain. The large tubes are isosurfaces corresponding
to $60\%$ of the maximum vorticity modulus. These tubes cross along the periodic $Y$-Axis and deform notably near the symmetry plane ($y=0$). The
elongated blobs in the interior of the tubes are isosurfaces corresponding to $90\%$ of the maximum vorticity. These isosurfaces are very
localized and flattened near the symmetry plane.

\begin{figure*}
\vspace{-35mm}
\includegraphics[width=8.0cm,angle=0]{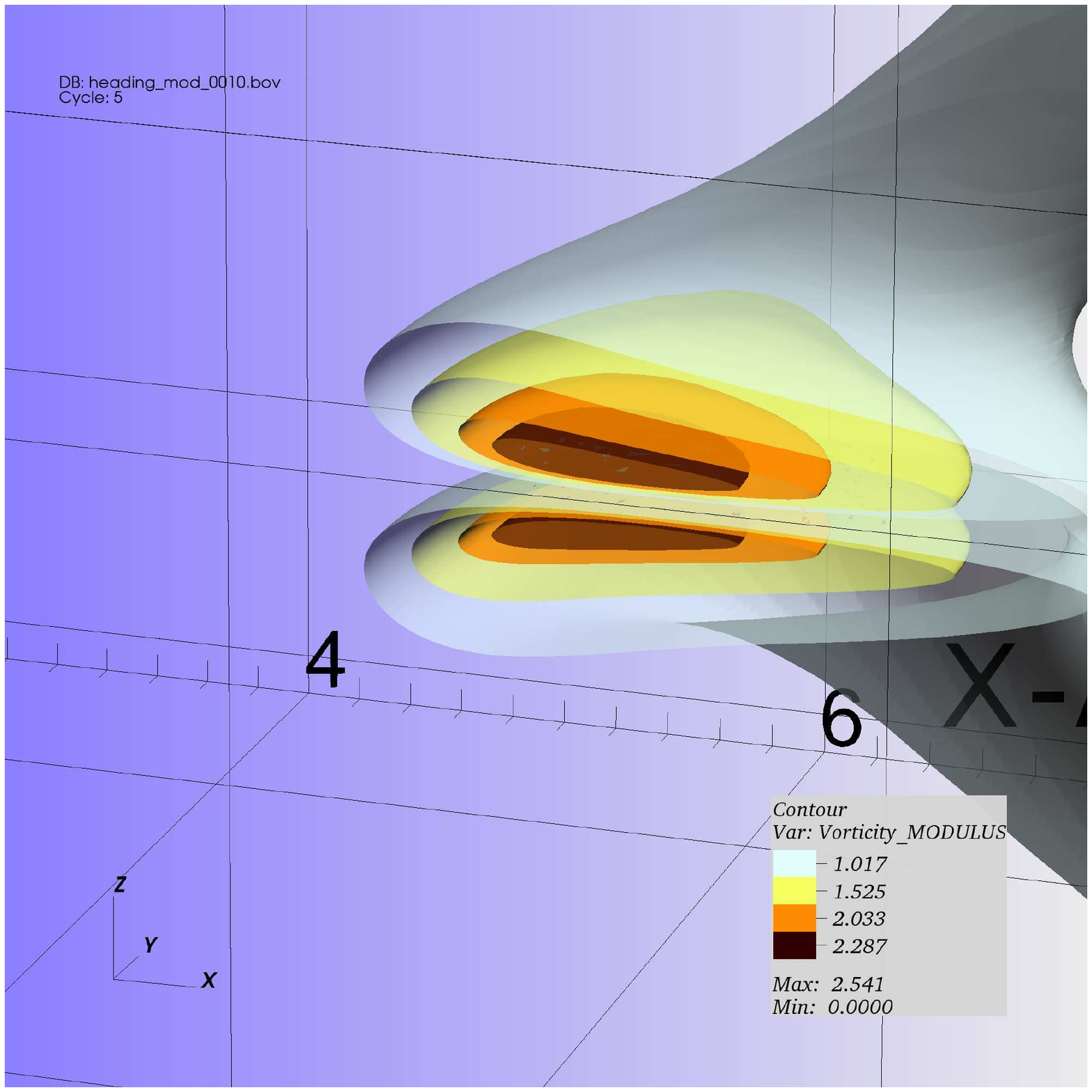}
\vspace{-35mm}
\includegraphics[width=8.0cm,angle=0]{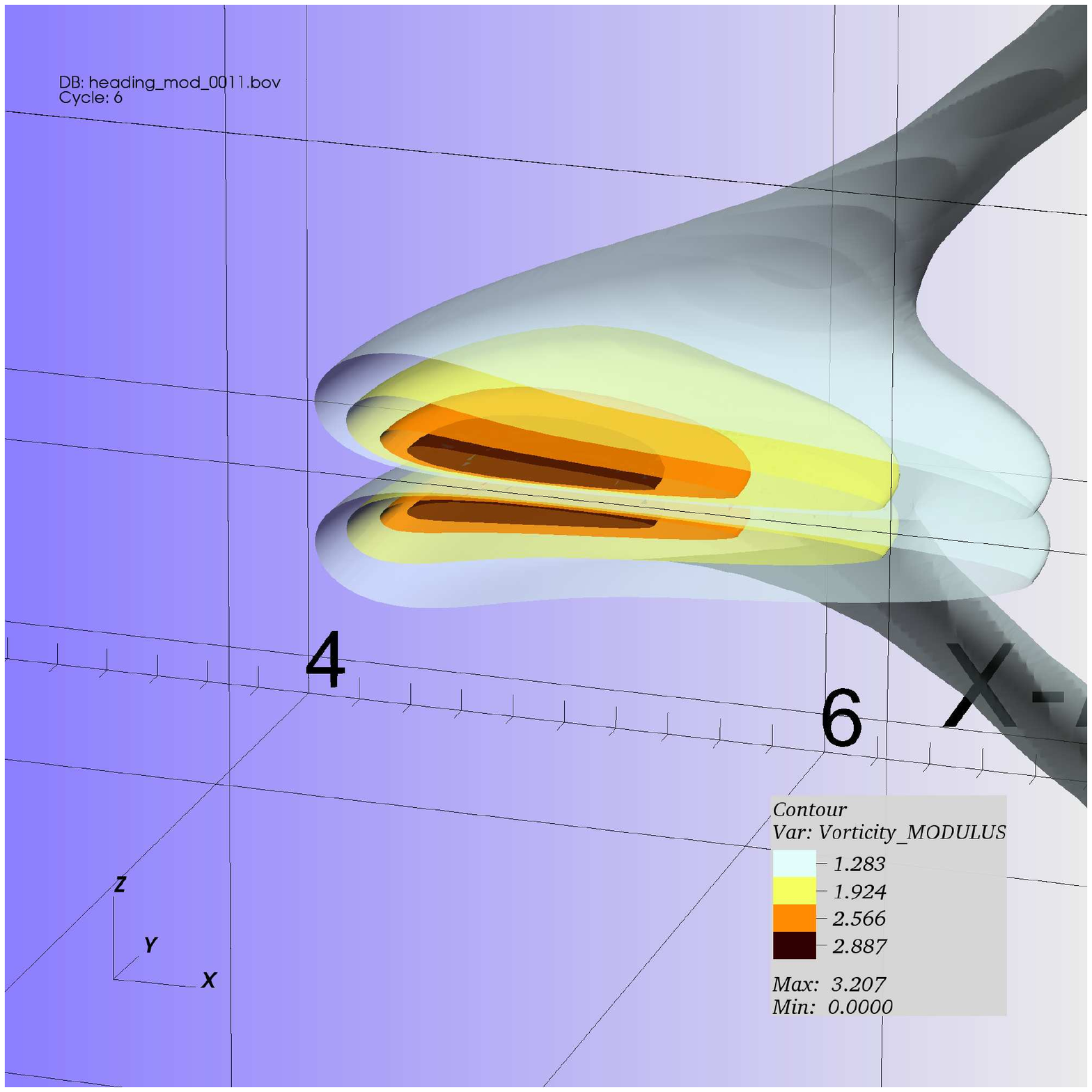}
\includegraphics[width=8.0cm,angle=0]{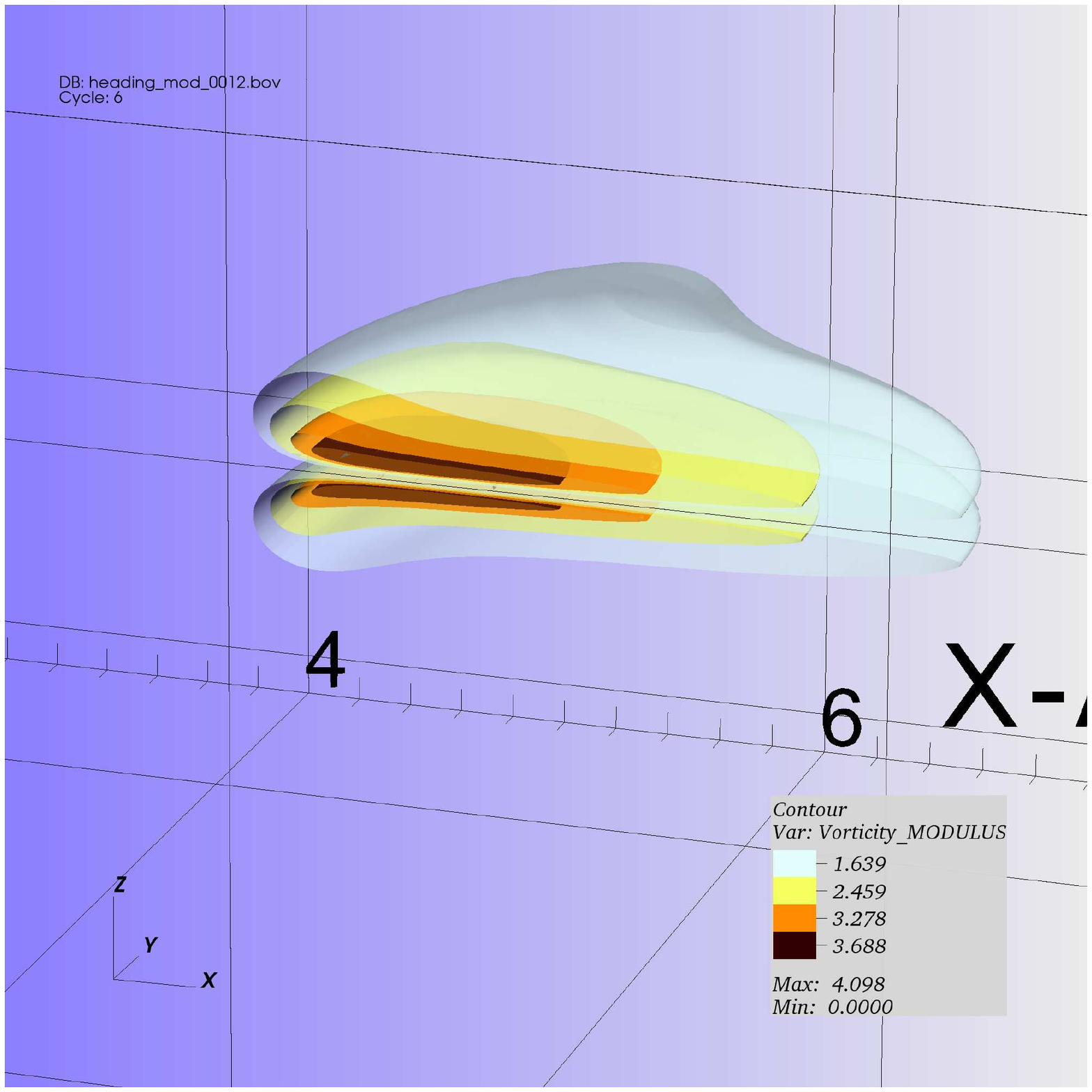}
\vspace{-35mm}
\includegraphics[width=8.0cm,angle=0]{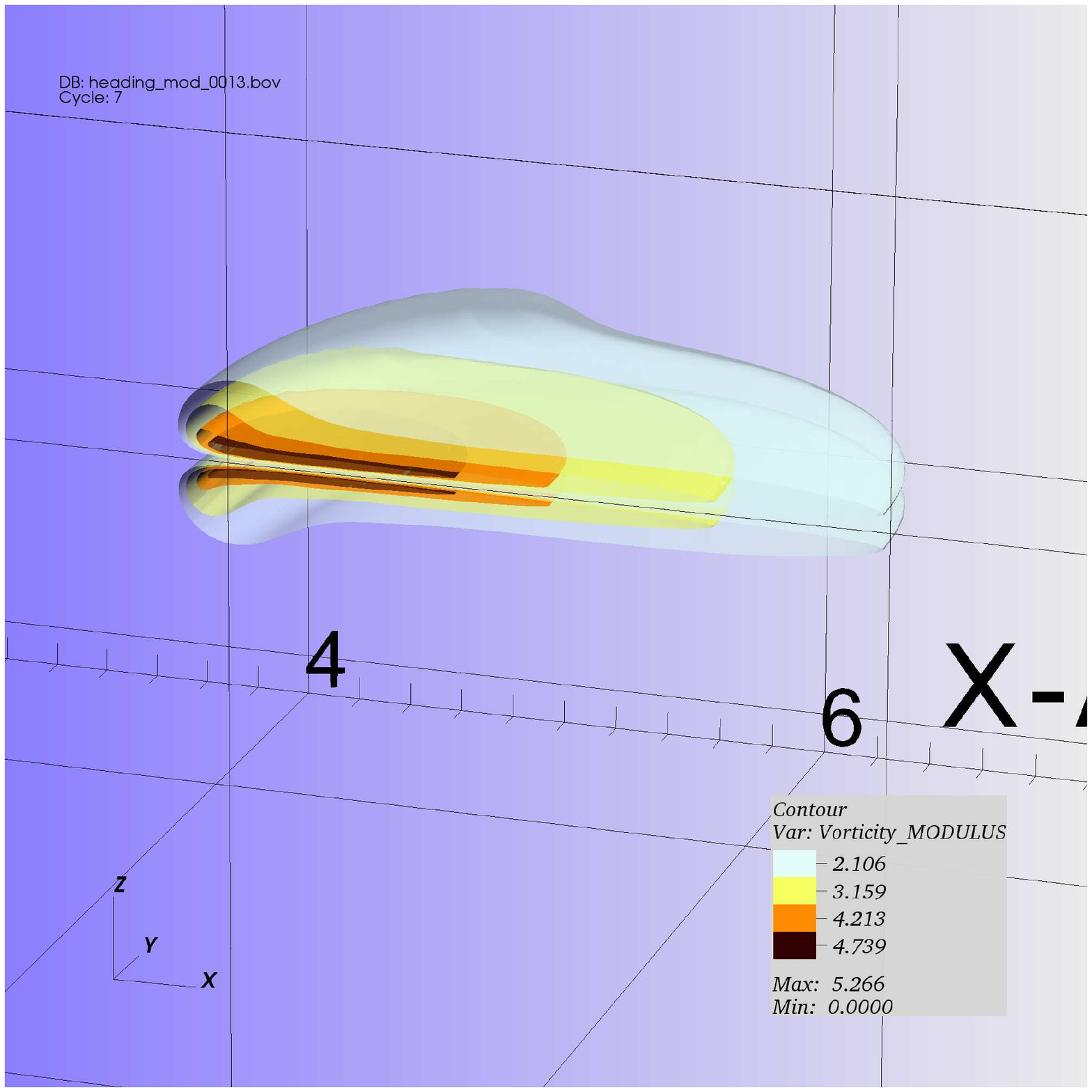}
\includegraphics[width=8.0cm,angle=0]{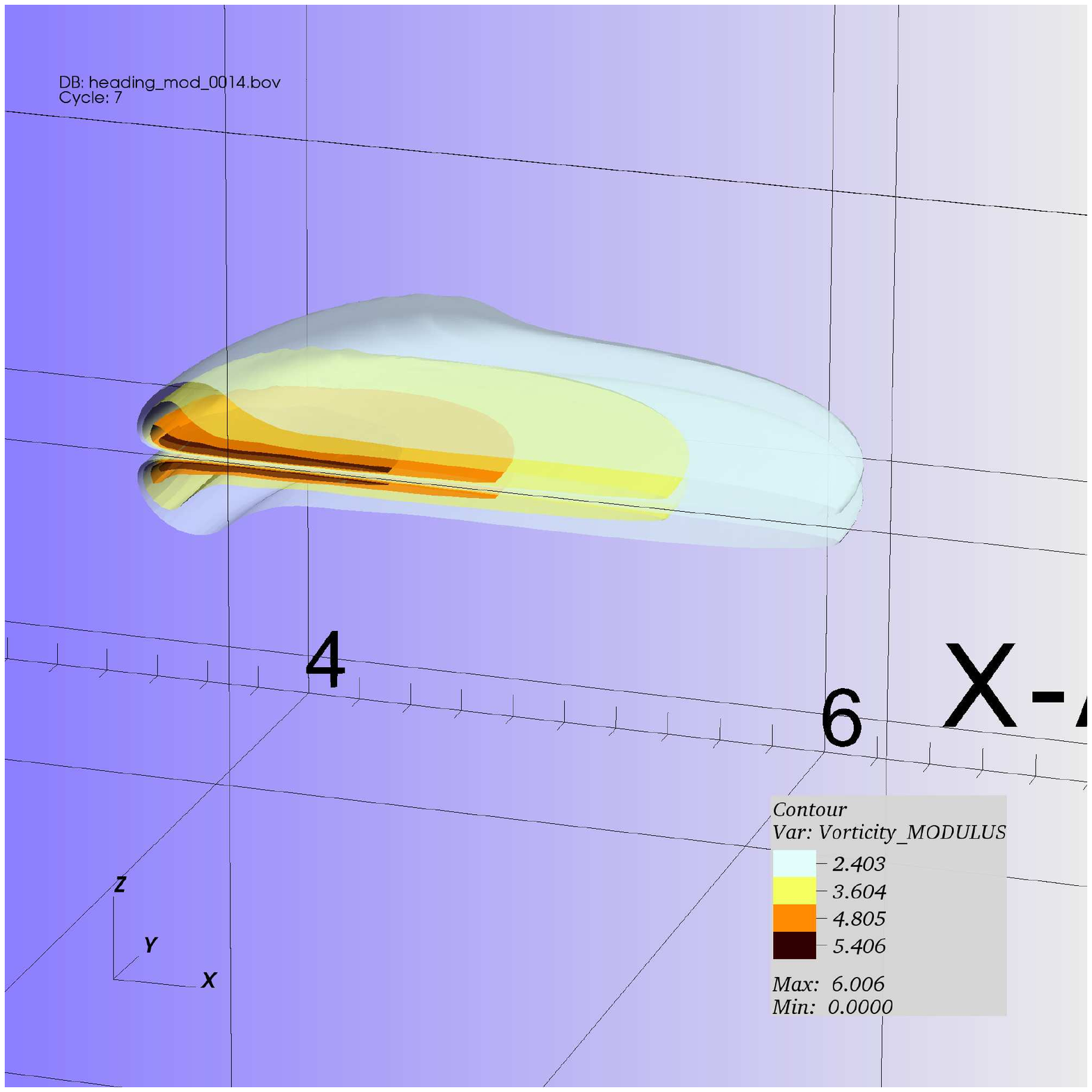}
\includegraphics[width=8.0cm,angle=0]{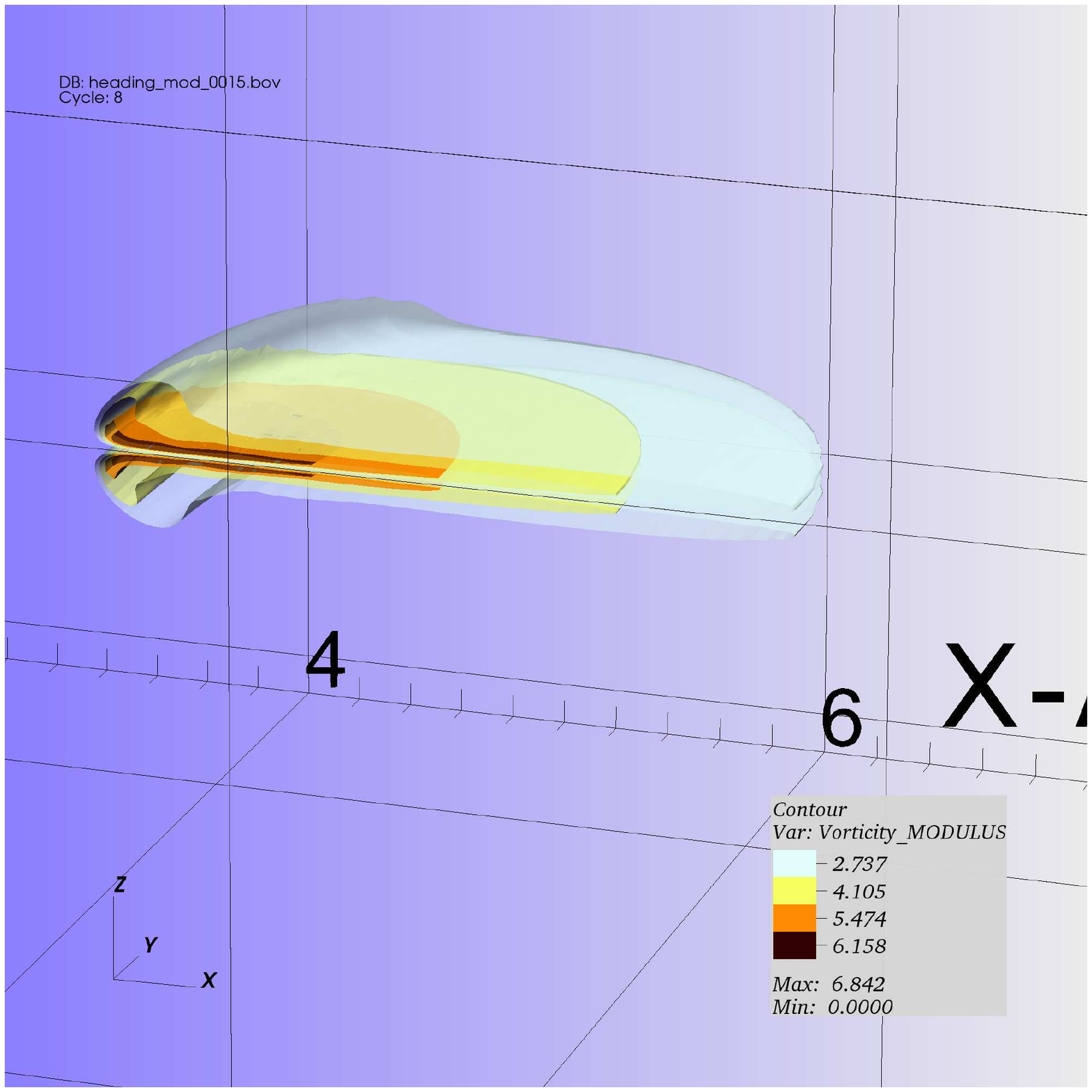}
\vspace{-0mm} \caption{From left to right, and from top to bottom: six successive, zoomed snapshots of the Euler anti-parallel vortices at times
$t=5.625, 6.25, 6.875, 7.5, 7.8125, 8.125.$ The contours are sectioned through the $y=0$ symmetry plane, to facilitate the view of the
structures. The contours are isosurfaces of vorticity modulus corresponding, respectively from outer to inner, to the $40\%, 60\%, 80\%$ and
$90\%$ of the value of the instantaneous maximum vorticity modulus.} \label{fig:v71snapshots}
\end{figure*}

Fig. \ref{fig:v71snapshots} shows successive snapshots at later stages of the flow, of isosurfaces corresponding to the following percentages of
the instantaneous value of the maximum vorticity modulus: from outer to inner contours, $40\%$, $60\%$, $80\%$ and $90\%$. Only half of the total
domain is shown so that a section of the isosurfaces through the symmetry plane ($y=0$) is visible. The snapshots are all seen from the same
angle and with the same zoom with respect to the fixed box. To read the snapshots going forward in time one advances from left to right and from
top to bottom.

The flattening in the $z$-direction results in structures similar to flattened pillows with some curvature in $x$ (and less in $y$) that becomes
more pronounced at the later times along the cut at the symmetry plane. These empirical observations provide further support for the choice of
anisotropic resolution in the simulations.


\section{Looking forward}

The calculation reported here was the largest possible on the Warwick SGI Altix with our code.  In the near future we will have a cluster capable
of simulating a $2048\times1024\times4096$ mesh. We might also use UK national computing resources.

Once our new cluster arrives we anticipate the following calculations: \ITM\item Further tests to determine if an initial condition closer to
that of Kerr (1993) \cite{Kerr93} can be obtained.
\item After further resolution checks, at least one calculation
on a $2048\times512\times4096$ mesh either on the new profile here or that of Kerr (1993) \cite{Kerr93}.
\item At least one modest resolution calculation on the square-off
initial condition of Fig. \ref{fig:omy}.
\item  Our goal in high resolution calculations will be to include spectral convergence tests, in particular the analyticity strip method
(Sulem et al. \cite{1983JCoPh..50..138S}, see also \cite{{Fri03},Cich04} and references therein) which gives independent evidence of
singular/nonsingular behavior of the flow and allows one to extrapolate the convergence of $\ominfty$.
\item Convergence of $\ominfty$ and other local quantities should allow us to study regularity
bounds from Constantin, Fefferman and Majda \cite{ConstFM96} and from Deng, Hou and Yu \cite{DHY06}. \ITN

The finite-time singularity hypothesis of the three-dimensional, incompressible Euler equations, leads to conclusions that are in qualitative
agreement with Kerr (1993) \cite{Kerr93}. However, we have found that the previously proposed scaling laws and estimated singular time must be
modified.

One possible outcome which will require further investigation is whether constant circulation is trapped within the collapsing region. If this is
confirmed, the two length scaling parameterization proposed \cite{Kerr05} cannot be correct.

It is anticipated that in addition to these much higher resolution anti-parallel calculations, there will soon be new high-resolution
calculations of the Kida-Pelz flow (anticipated in these proceedings by the contribution of Grafke et al. \cite{Gra07}) and the Taylor-Green
calculations initiated by Brachet et al. (1983) \cite{Brachetetal83} will soon be continued by Brachet.  If these prove to be
singular, and anti-parallel, it is possible that they will reproduce the scalings hinted at here.\\
\\




\begin{acknowledgments}

We acknowledge discussions with C. Bardos, J. D. Gibbon, R. Grauer, T. Hou, S. Kida, R. Morf, K. Ohkitani, E. Titi and other participants in
Euler250. We thank U. Frisch and co-workers for organizing this excellent symposium.  Support for this work was provided by the Leverhulme
Foundation grant F/00 215/AC. Computational support was provided by the Warwick Centre for Scientific Computing.
\end{acknowledgments}


\begin{thebibliography}{1}

\bibitem{Euler1761}
\authone{L.}{Euler}\yjour{1761}{Novi Commentarii Acad. Sci. Petropolitanae} {6}{271}{--311}{Principia motus fluidorum.}

\bibitem{BKM84} 
\auththr{J. T.}{Beale}{T.}{Kato}{A.}{Majda} \yjour{1984}{Commun. Math. Phys.}{94}{61}{} {Remarks on the breakdown of smooth solutions for the
$3D$ Euler equations.}

\bibitem{Kerr93} 
\authone{R.M.}{Kerr} \yjour{1993}{Phys. Fluids}{5}{1725}{-1746} {Evidence for a singularity of the three-dimensional, incompressible Euler
equations.}

\bibitem{ConstFM96} 
\auththr{P.}{Constantin}{C.}{Fefferman}{A.}{Majda} \yjour{1996}{Comm. Partial. Diff. Equns.}{21}{559}{-571} {Geometric constraints on potentially
singular solutions for the 3D Euler equations.}

\bibitem{Kerr05} 
\authone{R.M.}{Kerr}\yjour{2005}{Phys. Fluids}{17}{075103}{} {Velocity and scaling of collapsing Euler vortices.}

\bibitem{Graueretal98}\auththr{R.}{Grauer}{C.}{Marliani}{K.}{Germaschewski}
\yjour{1998}{Phys. Rev. Lett.}{80}{4177}{4180} {Adaptive mesh refinement of singular solutions of the incompressible Euler equations.}

\bibitem{OrlandiCarnevale07}\authtwo{P.}{Orlandi}{G.}{Carnevale}
\yjour{2007}{Phys. Fluids}{19}{057106}{} {Nonlinear amplification of vorticity in inviscid interaction of orthogonal Lamb dipoles.}

\bibitem{HouLi06} 
\authtwo{T.Y.}{Hou}{R.}{Li} \yjour{2006}{J. Nonlin. Sci.}{16}{639}{--664} {Dynamic depletion of vortex stretching and non-blowup of the 3-D
incompressible Euler equations.}



\bibitem{KerrH89} 
\authtwo{R.M.}{Kerr}{F.}{Hussain}\yjour{1989} {Physica D}{37}{474}{-484}{Simulation of vortex reconnection.}

\bibitem{Kerr92} 
\authone{R.M.}{Kerr} \yproc{1992}{309}{--336} {Topological aspects of the dynamics of fluids and plasmas} {G.M. Zaslavsky, M.  Tabor \biband P.
Comte} {Proceedings of the NATO-ARW workshop at the Institute for Theoretical Physics, University of California at Santa Barbara. Kluwer Academic
Publishers, Dordrecht, Netherlands.} {Evidence for a singularity of the three-dimensional incompressible Euler equations.}

\bibitem{HouLi07} 
\authtwo{T.Y.}{Hou}{R.}{Li} \yjour{2007}{J. Comp. Phys.}{226}{379}{--397} {Computing nearly singular solutions using pseudo-spectral methods.}

\bibitem{1983JCoPh..50..138S} \auththr{C.}{Sulem}{P.-L.}{Sulem}{H.}{Frisch}
\yjour{1983}{J. Comp. Phys.}{50}{138}{-161} {Tracing complex singularities with spectral methods.}


\bibitem{Cich04}
\authtwo{C.}{Cichowlas}{M.E.}{Brachet} \yjour{2004}{Fluid Dynamics Research}{36}{239}{-248} {Evolution of Complex Singularities in {K}ida-{P}elz
and {T}aylor-{G}reen Inviscid Flows.}

\bibitem{Fri03} \auththr{U.}{Frisch}{T.}{Matsumoto}{J.}{Bec}
\yjour{2003}{J. Stat. Phys.}{113}{761}{-781} {Singularities of Euler flow? Not out of the blue!}

\bibitem{DHY06}
\auththr{J.}{Deng}{T.Y.}{Hou}{X.}{Yu} \yjour{2006}{Commun. Partial Diff. Equns.}{31}{293}{-306} {Improved geometric condition for non-blowup of
the 3{D} incompressible {E}uler equation.}


\bibitem{Gra07}
\authmanythr{T.}{Grafke}{H.}{Homann}{J.}{Dreher} \authone{R.}{Grauer} {Numerical simulations of possible finite time singularities in the
incompressible Euler equations: comparison of numerical methods (2007), these Proceedings.}


\bibitem{Brachetetal83} 
\authmanythr{M.E.}{Brachet}{D.I.}{Meiron}{S. A.}{Orszag} \auththr{B. G.}{Nickel}{R.H.}{Morf}{U.}{Frisch} \yjour{1983}{J. Fluid
Mech.}{130}{411}{--452}{Small-scale structure of the Taylor-Green vortex.}

\end{thebibliography}

\end{document}